\documentstyle[12pt,a4]{article}

\begin{document}
\title{\bf Study of $\Delta(1232)$ isobar electroproduction at VEPP-2M
           $e^+e^-$ collider.}
\author{M.N.Achasov\thanks{ E-mail: achasov@inp.nsk.su,
                            FAX: +7(383-2)35-21-63},
        A.V.Bozhenok, A.D.Bukin, D.A.Bukin, S.V.Burdin, \\
	T.V.Dimova, S.I.Dolinsky, V.P.Druzhinin, M.S.Dubrovin, \\
	I.A.Gaponenko, V.B.Golubev, V.N.Ivanchenko, A.A.Korol, \\
	S.V.Koshuba, A.P.Lysenko, E.V.Pakhtusova, V.V.Shary, \\
	Yu.M.Shatunov, V.A.Sidorov, Z.K.Silagadze, S.I.Serednyakov, \\
	A.A.Valishev, Yu.V.Usov \\
\it     G.I.Budker Institute of Nuclear Physics, \\
\it     Siberian Branch of the Russian Academy of Sciences, \\
\it	630090,
        Novosibirsk,
	Russia }
			      
\date{}
\maketitle
\begin{abstract}
 Results from the Spherical Nonmagnetic Detector (SND) on $\Delta (1232)$
 isobar electroproduction in the collisions of beam electrons (positrons)
 and residual gas nuclei in the  VEPP-2M $e^+e^-$ collider are presented.
 On the basis of the obtained data the expected
 counting rate of this process in future high luminosity
 $e^+e^-$ colliders (~$\phi$-, $c$-$\tau$- and $b$-factories) was estimated.
\end{abstract}


 Electron or positron beam propagating along the collider beam pipe
 can produce pions on nuclei of residual gas atoms.
 At a beam energy of $E_0 \approx 500 \mbox{MeV}$ ($\phi$-factory region) the
 main source of such pions is an electroproduction of  $\Delta(1232)$
 $I(J^P)=\frac{3}{2}\left (\frac{3}{2}^+\right )$ isobar state
  \cite{ufn}.

 In an electron--nucleus collision a nucleon can convert
 into either $\Delta^+$ $(ep \rightarrow e \Delta^+)$ or $\Delta^0$
 $(en \rightarrow e \Delta^0)$ with a subsequent decay of $\Delta(1232)$
 into nucleon and pion: $\Delta^+ \rightarrow p \pi^0, n\pi^+$;
 $\Delta^0 \rightarrow p \pi^-, n\pi^0$. The total number of produced $\pi^0$
 is twice higher than that of $\pi^+$ or $\pi^-$.
 At an electron energy of about
 $500 \mbox{MeV}$ cross section of the isobar electroproduction is
 proportional to $A$, atomic weight of the target. But only those
 $\Delta$-isobars, which were produced on the nucleus surface, would
 actually decay into
 $\Delta \rightarrow N \pi$. For $\Delta$-s, produced inside the nucleus
 the $\Delta N \rightarrow NN$ reaction dominates.
 It means that the $e N \rightarrow e \Delta$, $\Delta \rightarrow \pi N$
 cross section must be proportional to $A^{2/3}$.

 The total cross section of the isobar electroproduction on a nucleon at
 electron energy of 510 MeV is about $3 \mu \mbox{b}$. This reaction is a
 possible source of background at $\phi$-factories \cite{Sech}. It can be also
 an actual background in experiments at higher energies
(its total cross section is about
 $8 \mu\mbox{b}$ at a $c-\tau$-factory and $12 \mu \mbox{b}$ in the $b$-factory
 energy region). This paper presents results of experimental study
 of $\Delta(1232)$ electroproduction with SND detector at VEPP-2M
 $e^+e^-$ collider in Novosibirsk.
 
 SND detector  \cite{snd} consists of a drift chamber tracking system
 with an angular resolution $\sigma_{\theta}=2.2^{\circ}$,
 $\sigma_{\phi}=0.7^{\circ}$
 and $dE/dx$ resolution of about $30 \%$, a three layer NaI(Tl) spherical
 electromagnetic calorimeter \cite{CCC} with an angular resolution
 $\sigma_{\theta}=\sigma_{\phi}=1.5^{\circ}$ and energy resolution 
 $\sigma_E/E=4.2\%/\sqrt[4]{E(\mbox{GeV})}$ for photons, and a muon
 identification system.

 The $\Delta$-isobar electroproduction  process was studied 
 in $ep \rightarrow e \Delta^+$,
 $\Delta^+ \rightarrow p \pi^0$ channel, because of its very distinct
 signature: two photons from $\pi^0$-meson decay and two tracks from
 electron and proton. It is important that in this case photon
 energies and invariant masses are known and electron/proton
 separation based on  $dE/dx$ measurements in the drift chambers is possible.
 
 Monte Carlo simulation of the $\Delta$-isobar electroproduction
 on a single proton, based on formulae from \cite{Sech}, was used
 for studying selection criteria
 and detection efficiency. Passage of particles through the 
 detector was simulated using UNIMOD2 code \cite{UNIMOD}.

 Experimental data collected in 1996 in the energy range
 $2E_0 = 1.00\div1.04 \mbox{GeV}$ \cite{hadron97}, with an integrated
 luminosity of $L=0.5 \mbox{pb}^{-1}$ were processed. Events with two
 charge particles and two photons were selected
 for further analysis. To suppress the background
 from decays of copiosly produced  $\phi$-mesons
 (~$\phi \rightarrow 3 \pi, K_SK_L, K^+K^-$~, etc.) the following selection
 criteria were applied: total energy deposition in the calorimeter
 is less than the beam energy $E_0$, the angle between charged particles is
 smaller than $150$ degrees, which greatly reduces
 $K^+K^-$ and $K_SK_L$ background,
 and $dE/dx$ of one of the charged particles is
 at least twice larger than that of a minimum ionizing particle.

 Coordinate of a $\Delta$-isobar production point along the beam was
 reconstructed using charged particles tracks. 
 Then the kinematic fit was performed under following constraints:
 the particle with a larger $dE/dx$ was considered as proton;
 total transverse momentum
 $p_\perp = 0$; longitudinal momentum $p_\parallel = E_0/c$, photons
 originate from $\pi^0 \rightarrow \gamma\gamma$ decay, and the
 total energy $E = E_0+m_pc^2$, where $m_p$ is a proton mass.
 
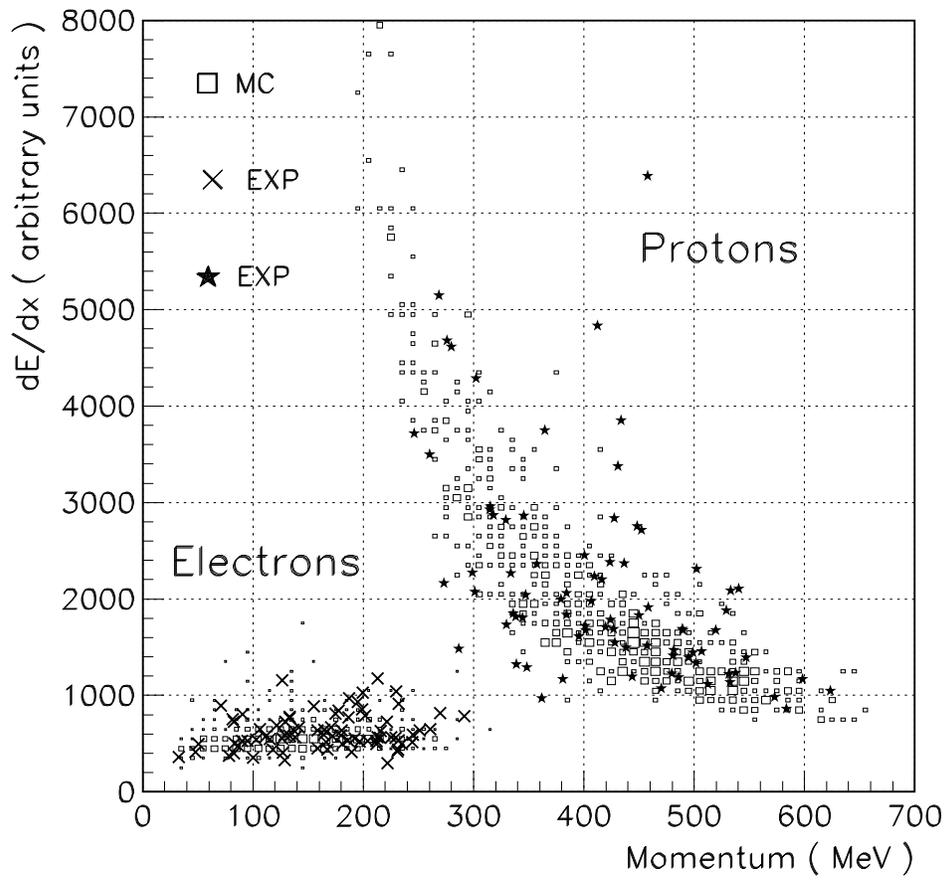
\begin{figure}
\caption{$dE/dx$, specific ionization losses of electrons and protons,
produced in $\Delta(1232)$ decays.
Crosses - electrons, stars - protons}
\label{fig23}
\end{figure}

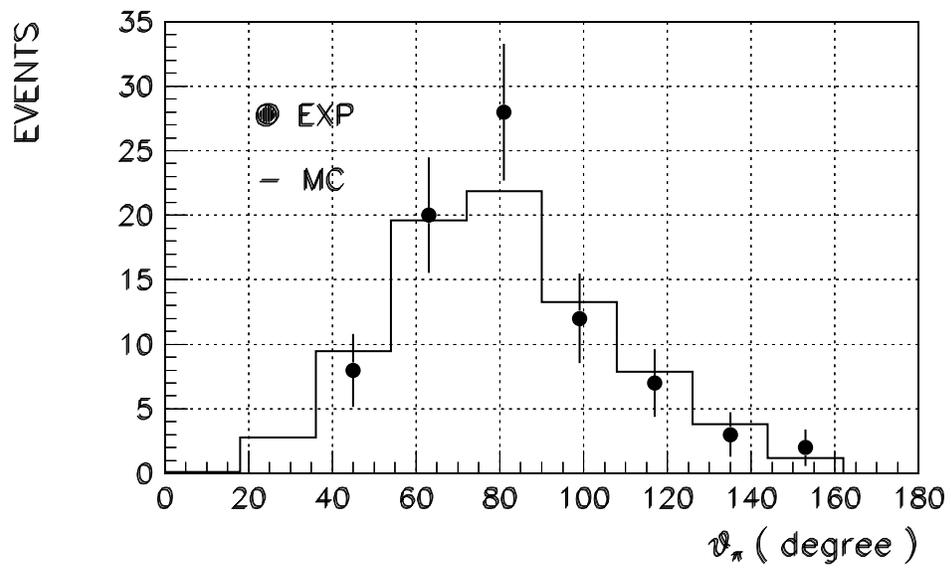
\begin{figure}
\caption{The $\pi^0$ polar angle distribution in $\Delta(1232)$ decay.}
\label{fig30}
\end{figure}

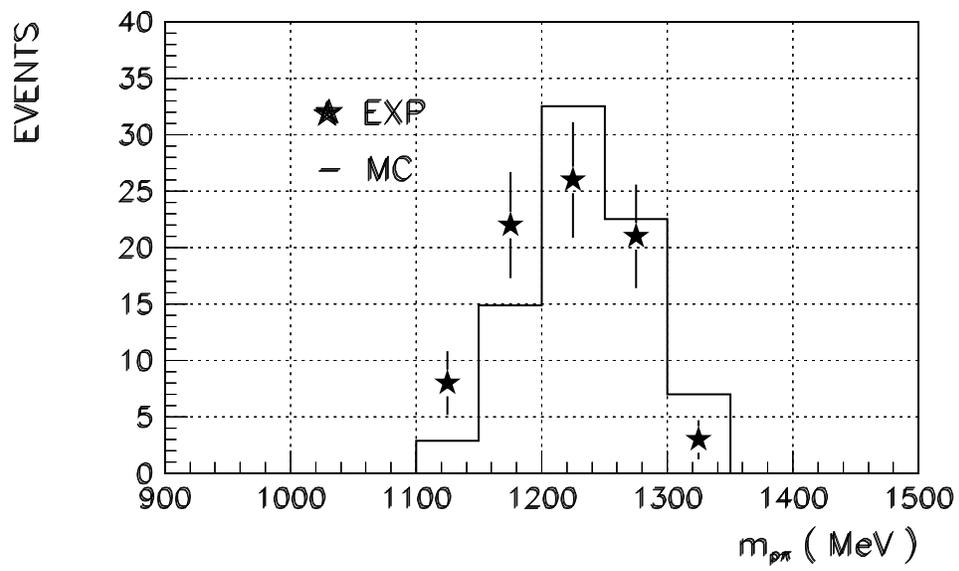
\begin{figure}
\caption{The $p\pi^0$ invariant mass distribution in the process
         of $\Delta(1232)$ electroproduction.}
\label{fig24}
\end{figure}

 Eighty events consistent with these  assumptions were found.
 Their production points are uniformly distributed along
 the beam direction within 20 cm fiducial length 
 in contrast with background, peaked at the beams collision point.
 The  dependence of $dE/dx$ on reconstructed charged
 particle momentum is shown
 in Fig.~\ref{fig23}. Electrons and protons are well separated and 
 momentum dependence of the proton specific ionization losses is
 clearly seen. The characteristic feature of the
 $\Delta \rightarrow \pi N$ decay is that  $\pi$-mesons are emitted
 at large angles with respect to the beam direction \cite{Sech}. 
 The experimental and simulated distributions, shown  in Fig.\ref{fig30}  are
 in a good agreement. The proton--pion invariant mass spectra
 (~fig.\ref{fig24}~) are  peaked between $1200 \div 1250 \mbox{MeV}$.
 The peaks are located at $1218 \pm 6 \mbox{MeV}$ and $1235 \pm 2 \mbox{MeV}$
 in experimental and simulated distributions respectively.

 The expected number of selected experimental events
 can be written as:
 \begin{eqnarray}
 N = { I \over e }
 \cdot t \cdot \sigma \cdot N_p \cdot l \cdot \epsilon,
 \end{eqnarray}
 where $I = 30 \mbox{mA}$ is an average beam current, $e$ -- electron charge,
 $t = 7\cdot10^5\mbox{ s}$ -- total data acquisition time,
 $l = 20 \mbox{cm}$ --
 fiducial length, $\sigma = 2 \mu \mbox{b}$, $N_p$ -- effective
 density of protons,
 $\epsilon = 0.026$ -- detection efficiency. Number of observed
 experimental events is $N = 80$ $\pm 9$ (statitical) $\pm 10$ (systematic).
 It corresponds to
 $N_p = 6 \cdot 10^{14}/\mbox{m}^3$. The residual gas pressure $P= nkT$,
 where $k$ is the Boltzmann constant, $T=300 \mbox{K}$ and $n$ is the
 density of residual gas molecules.
 The expected composition of the residual gas is
 $H_2$ -- $30 \%$, $CH_4$ -- $10\%$, $CO$ -- $20\%$ and $CO_2$ -- $40\%$.
 In this
 case $N_p = 6n$, and $P$, calculated using expression (1), is equal to
 $3 \pm 0.4 \mbox{(statistical)} \pm 2 \mbox{(systematic)~nTorr}$. 
 This value agrees with direct pressure measurements:
 $P=3 \pm 2 \mbox{nTorr}$. The
 large  systematic error in the former value is due to uncertainty in
 residual gas composition.

 At DA$\Phi$NE $\phi$-factory \cite{Vignola}
 $(I = 5 \mbox{A}, P=1~\mbox{nTorr})$ the rate of
 $ep \rightarrow e \Delta^+$, $\Delta^+ \rightarrow p \pi^0$ reaction
 is $\sim 1.2 \mbox{Hz/m}$. Taking into account $\Delta^+ \rightarrow n \pi^+$
 decay and $\Delta^0$ electroproduction on the neutrons the counting
 rate raises up to $4 \mbox{Hz/m}$. This value agrees with estimation in
 \cite{Sech}.

 $\phi$-factory experimental program includes $CP$-violation in
 kaon decays, rare decays of $\phi$-meson, two-photon processes and
 $e^+e^- \rightarrow \mbox{hadrons}$ annihilation at low energies
 \cite{hb}. The $\Delta(1232)$ electroproduction process could be
 considerable source of background, for instance, for two-photon processes
 (the counting rate for
 $\gamma \gamma \rightarrow \pi^0$ is expected to be $0.2 \mbox{Hz}$ )
 \cite{gg}. Its counting rate is also comparable with the rate
 of the $\phi$-meson rare decays  
 $\phi \rightarrow \gamma \eta^{'}, \gamma f_0(980), \gamma a_0(980)$
 $\leq 0.5 \mbox{Hz}$, or with $K_L \rightarrow \pi \pi$ decays
 $~\leq 2 \mbox{Hz}~$.

 On the other hand $\sim 4 \cdot 10^7$ events of
 $\Delta(1232)$ decays would be produced at the $\phi$-factory
 per effective year $(10^7 \mbox{s})$ per one meter of fiducial length.
 This means, that in addition to $e^+e^-$ physics, experiments in the field
 of nuclear physics, e.g. studies of collective effects in nuclei
 ( so-called $\Delta$--$h$ states \cite{PN}), are possible.

 This work was supported in part by the Russian Foundation of Basic
 Research, Grant No. 96-15-96327, 97-02-18561, and by International
 Scientific Foundation, Grant No RPX000 and RPX300.

\newpage

\end{document}